\documentclass{article}
\usepackage[margin=1in]{geometry}
\usepackage{parskip}
\usepackage{amsmath}
\usepackage{amssymb}
\usepackage{mathtools}
\usepackage{booktabs}
\usepackage{enumerate}
\usepackage{color}
\usepackage[hidelinks]{hyperref}
\usepackage{natbib}
\usepackage{multirow}
\usepackage{graphicx}
\usepackage{mathdots}
\usepackage{caption}
\usepackage{subcaption}

\title{On technical considerations of UCI-regulated velodrome track design}
\author{Theodore Stanoev}
\date{}
\emergencystretch=\maxdimen
\begin{document}
\maketitle
\begin{abstract}
\noindent%
A novel approach to velodrome design for UCI-regulated tracks is presented.
The mathematical model uses differential geometry to form a three-dimensional ruled surface.
The surface accounts for the safety zone, blue band, and track region, the latter of which is comprised of three types of segments: straight lines, the arcs of circles, and connecting transition curves.
Following a first-principles approach, the general expressions are derived from the Frenet-Serret relations, as a function of the banking and curvature profiles, lengths of curve segments, and turn radii of the bends.
Given the underdetermined nature of the design problem, particular solutions are obtained using a least-squares minimization of an objective function, within the framework of numerical optimization.
Computer renderings of two designs, a symmetric assembly of quadrants as well as an asymmetric one, are presented to demonstrate the versatility of the approach, which may be used to design velodrome tracks of any UCI Category and track geometry specification.

\end{abstract}
\section{Introduction}
\label{sec:Introduction}
A velodrome is a sporting arena built for track cycling.
The design features are regulated by the governing body of the sport, the International Cycling Union (UCI).
Broadly speaking, the track follows an oval shape with a banked surface that varies in inclination along its length.
The specific details regarding the track geometry are often withheld as they are considered confidential and proprietary, with only generic information about its surface being released to the general public, such as the type of wood used in the construction, the number of nails, and the range of banking angles. 
Consequently, researchers have created their own models, which, over the years, have increased in sophistication.

Prior to the turn of the millennium, researchers focused on mathematical models to examine aspects of road cycling~(e.g., \citet{OldsEtAl1995} and~\citet{MartinEtAl1998}).
Shortly thereafter, the focus shifted toward track cycling, albeit with several limiting assumptions: \citet{MartinEtAl2006} assume the circular track designed;~\citet{LukesEtAl2006} assume the track is composed of two straights connected by two semicircles.
The latter assumption provides a better approximation and, as such, became a common track design for some studies~(e.g., \citet{UnderwoodJermy2010,UnderwoodJermy2014}, \citet{CaddyEtAl2017}).
However, other researchers have modelled the track using theodolite measurements from a surveying tool~(e.g., \citet{ChengEtAl2011}, \citet{WozniakAniol2011}, \citet{FittonSymons2018}), and even one group modelled the bends of the track as half ellipses as opposed to semicircles~(\citet{BenhamEtAl2020}).
However, none account explicitly for the gradual transition between the entrance and exit of the bends.
The first to do so are~\citet{LukesEtAl2012}, who assume a linear change in curvature.
\citet{Solarczyk2020} recognized this apparent oversight in his overview of track design, but did not provide further developments. 
It was not until the COVID-19 pandemic that transition curves were reintroduced to track cycling models in two concurrent, yet independent, studies by~\citet{FitzgeraldEtAl2021} and~\citet{BosEtAl2022}.
The former developed a model for which the transition curve is either linear or sinusoidal. 
The latter considered linear curvature as well as the so-called Bloss curvature.
These studies comprise the state-of-the-art of velodrome track design. 

Of the previously mentioned studies, the common purpose is the demonstration of a model that accounts for the power required to overcome the forces opposing the motion of a cyclist.
As a consequence, simplifications are made to the track design in order to demonstrate clearly the features of their models.
For example, the ubiquitous assumption that the measuring line is confined to the horizontal plane, of which the bicycle wheel path is often set to follow.
However, since the measuring line is at a fixed offset distance from the inner edge of the track, which varies in inclination along its length, it is necessarily a three-dimensional curve.
Another assumption is that the velodrome is symmetric about its two axes, which is not a requirement.
By contrast, this article is dedicated entirely to track design.
The novel approach is rooted in differential geometry and its generality yields a fully customizable, UCI-compliant formulation.


We begin with a description of the technical regulations that are mandated by the UCI.
Then, we present the velodrome as a ruled surface and use the Frenet-Serret relations to derive general expressions for the segments that generate the surface.
Subsequently, we demonstrate the approach to obtain polynomial expressions of curvature for the transition curve using Hermite interpolation.
Afterward, we determine the system of equations that constitute a UCI-compliant track, with respect to banking and curvature profiles, respective lengths of curve segments, and the turn radii.
Finally, we present the solutions for a symmetric and asymmetric track design to demonstrate the versatility of the formulation.

\section{UCI regulations}
\label{sec:UCI}
In this section, we detail the pertinent regulations that restrict the velodrome surface design; the complete list of technical specifications are listed in~\citet[Chapter~VI]{UCI}.


Velodromes are categorized into one of four categories by a homologation process that determines the level of competition that may be organized at the arena.
The track geometry is constrained by the following five articles.
\begin{enumerate}
	\item
    3.6.067 ({\it form}): ``%
    the inner edge of the track shall consist of two curves connected by two parallel straight lines. 
    The entrance and exit of the bends shall be designed so that the transition is gradual.''
    In Section~\ref{sec:directrix}, we provide expressions for the inner-edge segments (a straight line for the straightaway, the arc of a circle for the bend, and a transition curve to connect them) as a function of their curvature.
    
    \item
    3.6.068: ``%
    the {\it length} of the track must lie between~133 metres and~500 metres inclusive.~$[\,\dots\,]$
    For the World Championships and the Olympic Games the length must be~250 metres.''
    
    \item
    3.6.069: ``the length of the track shall be measured~20 cm above the inner edge of the track (the upper edge of the blue band).''
    In Section~\ref{sec:length}, we provide integral equations to calculate the length of a lap, as determined by the measuring line. 
    
    \item
    3.6.070:  ``%
    the {\it width} of the track must be constant throughout its length.
    Tracks approved in categories 1 and 2 must have a minimum width of~7 metres.
    Other tracks must have a width proportional to its length of~5 metres minimum.''
    The width, $w$, and radius of the bends, $R$, for various track lengths for Category~1 and~2 tracks are shown in Table~\ref{tab:A3.6.095}.
    Although, in this article, we restrict our attention to velodromes of~$L=250\,{\rm m}$ length, the formulation is valid for tracks of any category and length. 
    
    \item
    3.6.073 ({\it profile}): ``%
    at any point on the track, a cross section of the track surface must present a straight line.''
	In Section~\ref{sec:director}, we define the director curve, for which the rulings are necessarily straight lines.
\end{enumerate}

\begin{table}
	\centering
	\begin{tabular}{cccccc}
		\cmidrule{1-6}
		Length of the track & $L$ & 250 m & 285.714 m & 333.33 m & 400 m \\
		Radius of bends & $R$ & 19--25 m & 22--28 m & 25--35 m & 28--50 m \\
		Width & $w$ & 7--8 m & 7--8 m & 7--9 m & 7--10 m \\
		\cmidrule{1-6}
	\end{tabular}
	\caption{Radius and width ranges for various track lengths~(adapted from~\citet[Article 3.6.095]{UCI})}
	\label{tab:A3.6.095}
\end{table}

These restrictions permit a freedom of track design.
If we consider the track as an assembly of four quadrants, it is permissible for the design features in each quadrant to vary, thus yielding an asymmetric design.
For example, the angle of the exit from the banking in the London~2012 Olympic velodrome is steeper than the entry so that the cyclist is, in effect, always riding downhill by being ``catapulted'' along the straights~(\citet{Douglas2010}).
As such, our analysis considers the individual segments that comprise each quadrant and is used to design both symmetric and asymmetric velodrome tracks. 

Along with the track, the surface is comprised of a blue band and safety zone region.
Article~3.6.071 specifies that ``%
a rideable area sky-blue in colour known as the {\it blue band} must be provided along the inside edge of the track.
The width of this band must be at least~10\% of the width of the track and its surface must have the same properties as of the track.''
Article~3.6.072 determines that ``%
immediately inside the blue band there shall be a prepared and marked {\it safety zone}.
The combined width of the blue band and the safety zone shall be at least~4 metres for tracks of~250 metres and over, and~2.5 metres for tracks shorter than~250 metres.''
In Section~\ref{sec:surfaceRegions}, we give the expressions for both regions as a function of the track geometry. 

Lastly, there are three longitudinal markings that are drawn on the track surface.
Article~3.6.079 states that``%
a line $[\,\dots\,]$ known as the {\it measuring line} shall be drawn at 20 cm from the inside edge of the track. 
The measurement of the measuring line shall be taken on its inside edge.''
The others~(Articles~3.6.080 and~3.6.081) are the red {\it sprinters' line} and blue {\it stayers' line}, whose respective inner edges are marked out at 85\,cm and $\max\{w/3,2.45\,{\rm m}\}$ distances.

\section{Ruled surface}
\label{sec:ruledSurface}
Following the nomenclature of~\citet[Definition~14.1]{GrayEtAl2006}, we define the velodrome as a ruled surface comprised of ruled patches of the form
\begin{equation}
	\label{eq:ruled_X(s,v)}
	{\bf X}(s,v) = {\boldsymbol\alpha}(s) + v\,{\boldsymbol\gamma}(s), 
\end{equation}
where $\boldsymbol\alpha$ is the directrix of the ruled surface, $\boldsymbol\gamma$ is the director curve, and the rulings are the straight lines~$v\mapsto{\boldsymbol\alpha}(s)+v\,{\boldsymbol\gamma}(s)$.

In this section, we derive general plane curve expressions as a function of curvature for the segments that comprise the directrix: a straight line, transition curve, and arc of a circle.
Next, we connect the segments to produce a simple, closed, and smooth directrix curve.
We proceed to determine the expression for the director curve so that the rulings generate straight-line cross sections of the surface. 
Then, we use the directrix and director to obtain the surface expression on the three regions that comprise the surface: the safety zone, blue band, and track.
Finally, we determine the expression to measure length on the surface to mark the three longitudinal markings of the track: the measuring, sprinters', and stayers' lines. 

\subsection{Frenet-Serret relations}
\label{sec:FS}
Let us derive general equations for the directrix segments as a function of their curvature, $\kappa(s)$.
The system of coordinates is Cartesian and is oriented to follow the right-hand rule convention.
We consider the Frenet-Serret relations for a differentiable curve in $\mathbb{R}^3$, whose matrix form is
\def\arraystretch{1}
\begin{equation}
	\label{eq:FS_3D}
	\left[\begin{array}{c}{\bf T}'(s) \\ {\bf N}'(s) \\ {\bf B}'(s)\end{array}\right]
	=
	\left[\begin{array}{ccc}0 & \kappa(s) & 0 \\ -\kappa(s) & 0 & \tau(s) \\ 0 & -\tau(s) & 0\end{array}\right]
	\left[\begin{array}{c}{\bf T}(s) \\ {\bf N}(s) \\ {\bf B}(s)\end{array}\right],
\end{equation}
where $\tau(s)$ is torsion, $(\circ)' = \frac{{\rm d}(\circ)}{{\rm d}s}$, and bold font indicates a vector quantity.
With respect to the position vector,~${\bf r}(s)$, the unit tangent, unit normal, and binormal vectors are
\begin{equation*}
	{\bf T}(s) = \frac{{\bf r}'(s)}{\|{\bf r}'(s)\|},\quad
	{\bf N}(s) = \frac{{\bf T}'(s)}{\|{\bf T}'(s)\|},\quad
	{\bf B}(s) = {\bf T}(s)\times{\bf N}(s),
\end{equation*}
respectively.
For plane curves, ${\bf B}(s)$ is perpendicular to the plane for all $s$ and, as such, we require ${\bf B}'(s) = 0$ and $\tau(s)=0$ to satisfy relations~\eqref{eq:FS_3D}.
Selecting the horizontal plane, the Frenet-Serret relations, expressed componentwise in $\mathbb{R}^2$, reduce to
\begin{equation*}
	\left[\begin{array}{cc} T_1'(s) & T_2'(s) \\ N_1'(s) & N_2'(s)\end{array}\right]
	=
	\left[\begin{array}{cc}0 & \kappa(s) \\ -\kappa(s) & 0\end{array}\right]
	\left[\begin{array}{cc} T_1(s) & T_2(s) \\ N_1(s) & N_2(s)\end{array}\right],
\end{equation*}
which form a two-dimensional system of homogeneous first-order linear ordinary differential equations with variable coefficients in the form
\begin{equation}
	\label{eq:ODE}
	Y'(s) = M(s)\,Y(s).
\end{equation}
According to~\citet[Section~5]{Magnus1954}, provided that system~\eqref{eq:ODE} is subject to initial conditions $Y(s_0) = I$, where $I$ is the identity matrix, the system always has a uniquely determined continuous solution, $Y(s)$, and has a continuous first derivative on any interval in which $M(s)$ is continuous.
Under the stipulation that continuity condition
\begin{equation}
	\label{eq:ODE_continuity}
	M(s)\,{\sf M}(s)\equiv {\sf M}(s)\,M(s)
\end{equation}
holds identically for all values $s$ within the interval, the matrix exponential $Y(s) = \exp{\sf M}(s)$ satisfies system~\eqref{eq:ODE}.
Since
\begin{equation*}
	{\sf M}(s) := \int_{s_0}^s M(\sigma)\,{\rm d}\sigma =  \left[\begin{array}{cc} 0 & \int_{s_0}^s\kappa(\sigma)\,{\rm d}\sigma \\ -\int_{s_0}^s\kappa(\sigma)\,{\rm d}\sigma & 0\end{array}\right] =: \left[\begin{array}{cc}0 & k(s) \\ -k(s) & 0\end{array}\right],
\end{equation*}
it is straightforward to demonstrate that the product of $M(s)$ and ${\sf M}(s)$ commutes.
To evaluate the matrix exponential, we avail of eigendecomposition property 
\begin{equation}
	\label{eq:MatExp}
	\exp({\sf M}) = {\sf Q}\exp({\sf\Lambda})\,{\sf Q}^{-1},
\end{equation}
where $\sf\Lambda$ is a diagonal matrix with the eigenvalues of $\sf M$ along its diagonal and $\sf Q$ is a matrix whose columns are the eigenvectors. 
For the decomposition, we solve the eigenvalue problem, ${\sf M}\,{\bf v} = \lambda\,{\bf v}$, where $\lambda$ is an eigenvalue of $\sf M$ and $\bf v$ is any nonzero vector. 
Proceeding in the usual way, the characteristic equation is
\begin{equation*}
	0 = \det({\sf M}-\lambda\,I) =
	\det\left[\begin{array}{cc}-\lambda & k(s) \\ -k(s) & -\lambda\end{array}\right] = \lambda^2 + k(s),
\end{equation*}
which implies that the two sets of eigenvalues and eigenvectors are $\lambda_1 = i\,k(s)$ and ${\bf v}_1 = [\,i,1\,]^t$, and $\lambda_2 = -i\,k(s)$ and ${\bf v}_1 = [\,-i,1\,]^t$, where ${}^t$ denotes the transpose.
Then,
\begin{equation*}
	{\sf\Lambda} = \left[\begin{array}{cc} i\,k(s) & 0 \\ 0 & -i\,k(s)\end{array}\right],\quad
	{\sf Q} = \left[\begin{array}{rc} -i & i \\ 1 & 1\end{array}\right],\quad
	{\sf Q}^{-1} = \frac{1}{2\,i}\left[\begin{array}{rc}-1 & i \\ 1 & i\end{array}\right] = \frac{1}{2}\left[\begin{array}{rc}i & 1 \\ -i & 1\end{array}\right],
\end{equation*}
which satisfies eigendecomposition ${\sf M} = {\sf Q}\,{\sf\Lambda}\,{\sf Q}^{-1}$.
Returning to matrix exponential~\eqref{eq:MatExp}, with the aid of trigonometric identities, we simplify to obtain a general solution to system~\eqref{eq:ODE}, where
\begin{equation}
	\label{eq:ODE_sol}
	Y(s) = \exp({\sf M}(s)) = \left[\begin{array}{rc}\cos k(s) & \sin k(s) \\ -\sin k(s) & \cos k(s)\end{array}\right].
\end{equation}
It is straightforward to verify that $Y'(s)=M(s)\,Y(s)$, which confirms matrix exponential~\eqref{eq:ODE_sol} as a solution to system~\eqref{eq:ODE}.
The initial conditions are verified by evaluating $k(s_0)$, which yields $Y(s_0)=I$, as required.
Also, without loss of generality, we recognize that, due to the periodicity of the trigonometric functions, solution~\eqref{eq:ODE_sol} holds under a phase shift by angle $\theta$. 
Thus, 
\begin{equation}
	\label{eq:ODE_sol_theta}
	\tilde{Y}(\theta,s) := \left[\begin{array}{rc}\cos(\theta + k(s)) & \sin(\theta + k(s)) \\ -\sin(\theta + k(s)) & \cos(\theta + k(s))\end{array}\right]
\end{equation}
is also a solution. 
\subsection{Directrix}
\label{sec:directrix}
The directrix design determines the geometry of the track.
Along the entrances of the bends, the curvature increases from~$\kappa\equiv0$ on the straightaways to~$\kappa\equiv1/R$ on the circular turns, by means of a transition curve.
Along the exits, the curvature decreases in an opposite manner. 
Since there are two such bends on a velodrome---as the entrance and exit of each bend consists of three curvature intervals---the directrix consists of twelve segments.

Although the directrix is in $\mathbb{R}^3$, it is confined to the $xy$-plane and, for convenience, we omit the $z$-component in this section.
The curve is simple, closed, and smooth, which requires that it never intersects itself, the initial and final coordinates are equal, and the tangent, as well as normal, vectors are equal at the beginning and end of each segment interval.
For our purposes, the positive direction corresponds to increasing arc-length values that trace out the curve in a counter-clockwise direction, emanating from an initial value~$s_0 := 0$. 
Using segment lengths,~${\boldsymbol l}$, the~$i$th interval is
\begin{equation*}
	S_i = [s_{i-1},s_i],
	\quad\text{where}\quad
	s_i = \sum\limits_{j=1}^il_j
	\quad\text{for}\quad
	i\in\{1,12\}.
\end{equation*}
Let us determine the tangent, normal, and position vector expressions for each segment.
We use the components of general solution~\eqref{eq:ODE_sol_theta} to define 
\def\arraystretch{1.25}%
\begin{equation*}
	{\bf T}_i(\theta_i,s) := \left[\begin{array}{c} \cos\left(\theta_i + \int_{s_{i-1}}^s\kappa(\varsigma)\,{\rm d}\varsigma\right) \\ \sin\left(\theta_i+ \int_{s_{i-1}}^s\kappa(\varsigma)\,{\rm d}\varsigma\right)\end{array}\right]
	\quad\text{and}\quad
	{\bf N}_i(\theta_i,s) = \left[\begin{array}{r} -\sin\left(\theta_i + \int_{s_{i-1}}^s\kappa(\varsigma)\,{\rm d}\varsigma\right) \\ \cos\left(\theta_i + \int_{s_{i-1}}^s\kappa(\varsigma)\,{\rm d}\varsigma\right)\end{array}\right],
	\quad\text{for}\quad s\in S_i.
\end{equation*}
Since we use an arc length parameterization, $\|{\bf r}'(s)\|=1$.
Then, we use separation of variables to write~${\rm d}{\bf r}_i(s) = {\bf T}_i(s)\,{\rm d}s$.
By the fundamental theorem of calculus, integrating both sides results in
\begin{equation*}
	{\bf r}_i(\theta_i,s) := ({\bf r}_0)_i + \int_{s_{i-1}}^s\left[\begin{array}{c} \cos\left(\theta_i + \int_{s_{i-1}}^\sigma\kappa(\varsigma)\,{\rm d}\varsigma\right) \\ \sin\left(\theta_i + \int_{s_{i-1}}^\sigma\kappa(\varsigma)\,{\rm d}\varsigma\right)\end{array}\right]{\rm d}\sigma,
\end{equation*}
\def\arraystretch{1}%
which determines the position vector along the~$i$th segment,~$S_i$, up to initial coordinates,~$({\bf r}_0)_i$, phase angles,~$\theta_i$, and curvature,~$\kappa_i(s)$.

The following three conditions ensure the geometric properties of the directrix.
First, for a smooth curve,
\begin{equation}
	\label{eq:theta_{i+1}}
	{\bf T}_i(\theta_i,s_i) = {\bf T}_{i+1}(\theta_{i+1},s_i)
	\implies
	\theta_i + \int_{s_{i-1}}^{s_i}\kappa_i(\varsigma)\,{\rm d}\varsigma = \theta_{i+1},
\end{equation}
which is recurrence relation that determines the~$(i+1)$th phase angle in terms of the~$i$th phase angle and curvature integration. 
Also, relation~\eqref{eq:theta_{i+1}} satisfies~${\bf N}_i(\theta_i,s_i) = {\bf N}_{i+1}(\theta_{i+1},s_i)$.
Similarly, ${\bf r}_i(\theta_i,s_i) = {\bf r}_{i+1}(\theta_{i+1},s_i)$ implies
\begin{equation*}
	({\bf r}_0)_i + \int_{s_{i-1}}^{s_i}\left[\begin{array}{c} \cos\left(\theta_i + \int_{s_{i-1}}^\sigma\kappa_i(\varsigma)\,{\rm d}\varsigma\right) \\ \sin\left(\theta_i + \int_{s_{i-1}}^\sigma\kappa_i(\varsigma)\,{\rm d}\varsigma\right)\end{array}\right]{\rm d}\sigma = ({\bf r}_0)_{i+1},
\end{equation*}
which determines that the final coordinates of the $i$th segment are the initial coordinates of the $(i+1)$th segment.
Second, for a closed curve, 
\begin{equation*}
	{\bf T}_{12}(\theta_{12},s_{12}) = {\bf T}_1(\theta_1,s_1)
	\quad\text{and}\quad
	{\bf N}_{12}(\theta_{12},s_{12}) = {\bf N}_1(\theta_1,s_1)
\end{equation*}
require
\begin{equation}
	\label{eq:theta_{12}}
	\theta_{12} + \int_{s_{11}}^{s_{12}}\kappa_{12}(\varsigma)\,{\rm d}\varsigma = \theta_{1} + 2\pi.
\end{equation}
Herein, $s_{12}$ is the length of one complete arc length cycle or, in other words, the period of the track. 
Given the periodicity of the cosine and sine functions, we include $2\pi$ to maintain their equality.
Also, ${\bf r}_{12}(\theta_{12},s_{12}) = {\bf r}_1(\theta_1,s_1)$ implies
\begin{equation}
	\label{eq:r_{12}}
	({\bf r}_0)_{12} + \int_{s_{11}}^{s_{12}}\left[\begin{array}{c} \cos\left(\theta_{12} + \int_{s_{11}}^\sigma\kappa_{12}(\varsigma)\,{\rm d}\varsigma\right) \\ \sin\left(\theta_{12} + \int_{s_{11}}^\sigma\kappa_{12}(\varsigma)\,{\rm d}\varsigma\right)\end{array}\right]{\rm d}\sigma = ({\bf r}_0)_{1}
\end{equation}
Third, for a simple curve, given that it is already smooth and closed, we require~$\kappa_i\geq0$ for each~$s\in S_i$.
Finally, the position, tangent, and normal vectors of the directrix are
\begin{equation*}
	{\boldsymbol\alpha}(s) := \sum\limits_{i=1}^{12}{\bf 1}_i(s)\,{\bf r}_i(\theta_i,s),\quad
	{\boldsymbol T}(s) := \sum\limits_{i=1}^{12}{\bf 1}_i(s)\,{\bf T}_i(\theta_i,s),\quad
	{\boldsymbol N}(s) := \sum\limits_{i=1}^{12}{\bf 1}_i(s)\,{\bf N}_i(\theta_i,s),
\end{equation*}
where the indicator function,
\begin{equation*}
	{\bf1}_i(s)
	:=
	\begin{dcases}
		1\quad\text{if}\quad s\in S_i \\[-2.5pt]
		0\quad\text{if}\quad s\not\in S_i
	\end{dcases},
\end{equation*}
serves as a toggle switch for the intervals based on arc length.

\subsection{Director curve}
\label{sec:director}
The director curve is used to generate the ruled surface. 
To determine its form, let us consider the distance between~${\boldsymbol\alpha}(s_0)$ and
\begin{equation*}
	P(s_0,\epsilon) = {\boldsymbol\alpha}(s_0) + \epsilon\left[\begin{array}{c}-({\boldsymbol N}(s_0))_x\\-({\boldsymbol N}(s_0))_y\\\tan\phi(s_0)\end{array}\right],
\end{equation*}
where~$\epsilon$ is a small distance in the outward normal direction along a slope of inclination~$\phi(s_0)$. 
Herein, $(\circ)_x$ and $(\circ)_y$ indicate the $x$- and $y$-components of $(\circ)$.
The distance between the two points is 
\begin{equation*}
	v = \sqrt{\left(P(s_0,\epsilon)-{\boldsymbol\alpha}(s_0)\right)\cdot\left(P(s_0,\epsilon)-{\boldsymbol\alpha}(s_0)\right)}
	= \epsilon\sqrt{({\boldsymbol N}(s_0))_x^2+({\boldsymbol N}(s_0))_y^2 + \tan^2\phi(s_0)}.
\end{equation*}
Solving for $\epsilon$ and substituting in the expression for the second point, we have
\begin{equation*}
	P(s_0,\epsilon) = {\boldsymbol\alpha}(s_0) + \frac{v}{\sqrt{({\boldsymbol N}(s_0))_x^2+({\boldsymbol N}(s_0))_y^2 + \tan^2\phi(s_0)}}\left[\begin{array}{c}-({\boldsymbol N}(s_0))_x\\-({\boldsymbol N}(s_0))_y\\\tan\phi(s_0)\end{array}\right].
\end{equation*}
Since this ruling is valid for each point along the directrix,  moving this straight line along the tangent vector of the curve generates a ruled surface ${\bf X}(s,v)$, where we define the director curve as
\begin{equation*}
	{\boldsymbol\gamma}(s) := \frac{1}{\sqrt{({\boldsymbol N}(s))_x^2+({\boldsymbol N}(s))_y^2 + \tan^2\phi(s)}}\left[\begin{array}{c}-({\boldsymbol N}(s))_x\\-({\boldsymbol N}(s))_y\\\tan\phi(s)\end{array}\right].
\end{equation*}
Since the banking inclination is restricted to $\phi(s)\in[0,\pi/2)$ and $\|{\boldsymbol N}(s)\|=1$ for each segment, trigonometric relation $\sqrt{({\boldsymbol N}(s))_x^2+({\boldsymbol N}(s))_y^2 + \tan^2\phi(s)} = \sec\phi(s)$ holds for all possible $\phi(s)$.
Also, since ${\boldsymbol N}(s)$ is confined to the $xy$-plane and ${\bf B}(s) = [0,0,1]^t$ for all $s$, the surface simplifies to
\begin{equation}
	\label{eq:surface}
	{\bf X}(s,v) = {\boldsymbol\alpha}(s) - v\left(\cos\phi(s)\,{\boldsymbol N}(s) - \sin\phi(s)\,{\bf B}(s)\right).
\end{equation}
\subsection{Surface regions}
\label{sec:surfaceRegions}
The velodrome surface is comprised of three regions: the safety zone, blue band, and track.
We define their respective widths as $w_{SZ}$, $w_{B}$, and $w$.
We set the origin of the ruling between the blue band and track, but the directrix is between the safety zone and blue band. 
To accommodate this configuration, we set the partitioning of the ruling domain as $v\in[-(w_{SZ}+w_B),-w_B]$ for the safety zone, $v\in[-w_B,0]$ for the blue band, and $v\in[0,w]$ for the track.
Furthermore, the inclination of each segment is as follows: $\phi=0$ for the safety zone, as it is in the horizontal plane; $\phi_B$ for the blue band, which is held constant along the length of the track; $\phi=\phi(s)$ for the track, as it varies with arc length.
Thus, we define the velodrome surface as
\begin{equation}
	\label{eq:VelodromeSurface}
	{\bf X}_V(s,v) := 
	\begin{dcases}
		{\boldsymbol\alpha}(s) - (v+w_B)\,{\boldsymbol N}(s), & v\in[-(w_{SZ}+w_B),-w_B] \\
		{\boldsymbol\alpha}(s) - (v+w_B)\left(\cos\phi_B\,{\boldsymbol N}(s) - \sin\phi_B\,{\bf B}(s)\right), & v\in[-w_B,0] \\
		{\boldsymbol\alpha}(s) - A(s,v)\,{\boldsymbol N}(s) +  B(s,v)\,{\bf B}(s), & v\in[0,w]
	\end{dcases},
\end{equation}
where
\begin{equation*}
	A(s,v) := w_B\cos\phi_B + v\cos\phi(s)
	\quad\text{and}\quad
	B(s,v) := w_B\sin\phi_B + v\sin\phi(s).
\end{equation*}

\subsection{Length}
\label{sec:length}
On the velodrome, the three longitudinal markings are drawn on the track, the measuring, sprinters', and stayers' lines, are drawn at fixed rulings \mbox{$v_M := 0.2$\,m}, \mbox{$v_{Sp} := 0.85$\,m}, and \mbox{$v_{St} := \max\{w/3,\,2.45\,{\rm m}\}$}, respectively.
Using expression~\eqref{eq:VelodromeSurface}, their coordinates are \mbox{${\bf r}_M(s) := {\bf X}_V(s,v_M)$}, \mbox{${\bf r}_{Sp}(s) := {\bf X}_V(s,v_{Sp})$}, and \mbox{${\bf r}_{St}(s) := {\bf X}_V(s,v_{St})$}.
Since the procedure to obtain the length is identical for any fixed ruling, we focus on the measuring line.

Following standard convention~(e.g.,~\citet[Definition~1.12]{GrayEtAl2006}), the length of a curve ${\bf c}(s)$ over an interval~$[s_0,s]$ is
\begin{equation*}
	L_{\bf c}(s_0,s) := \int_{s_0}^s\|{\bf c}'(\sigma)\|\,{\rm d}\sigma.
\end{equation*}
By the properties of the integral, integration over an interval comprised of subintervals is tantamount to the summation of the integration of each subinterval.
Thus, the length of a lap, as measured at ruling $v_M$, is 
\begin{equation}
	\label{eq:L_L}
	L_L 
	:= 
	L_{{\bf r}_M}(s_0,s_{12})
	=
	\sum\limits_{j=1}^{12}\left(\int_{s_{j-1}}^{s_j}\|{\bf r}_M'(\sigma)\|\,{\rm d}\sigma\right).
\end{equation}
To obtain the expression for $L_L$ along its intervals, we avail of the chain rule and Frenet-Serret relations~\eqref{eq:FS_3D} to write
\begin{equation*}
	{\bf r}_M'(s) 
	= 
	{\boldsymbol\alpha}'(s) + \kappa(s)A(s)\,{\boldsymbol T}(s) - A'(s){\boldsymbol N}(s) + B'(s){\bf B}(s),
\end{equation*}
where
\begin{equation*}
	A'(s) := \left.\frac{\partial A(s,v)}{\partial s}\right\vert_{v=v_M} = -v_M\sin\phi(s)\,\phi'(s)
    \quad\text{and}\quad
	B'(s) := \left.\frac{\partial B(s,v)}{\partial s}\right\vert_{v=v_M} = v_M\cos\phi(s)\,\phi'(s).
\end{equation*}
Using vector algebra,
\begin{equation*}
	\|{\bf r}_M'(s)\|
	=
	\sqrt{{\bf r}_M'(s)\cdot{\bf r}_M'(s)}
	=
	\sqrt{\left(1 + \kappa(s)\,A(s)\right)^2 + A'(s)^2 + B'(s)^2}.
\end{equation*}
Then, for example, with respect to the banking angle, blue band, and directrix intervals, the magnitude of the measuring line in the fourth quadrant is
\begin{equation*}
	\label{eq:r_M'(s)_mag}
	\|{\bf r}_M'(s)\| =
	\begin{dcases}
		\sqrt{1 + v_M^2\,\phi'(s)^2}, & s\in S_1 \\
		\sqrt{\left(1 + \kappa(s)\left(w_B\cos\phi_B + v_M\cos\phi(s)\right)\right)^2 + v_M^2\,\phi'(s)^2}, & s\in S_2 \\
		\sqrt{\left(1 + \frac{w_B\cos\phi_B + v_M\cos\phi(s)}{R_R}\right)^2 + v_M^2\,\phi'(s)^2}, & s\in S_3
	\end{dcases}
	\quad.
\end{equation*}

\section{Polynomial curvature}
\label{sec:CurvatureProfiles}
In this section, we detail transition curves, which we use to establish smooth connections between two segments with different magnitudes of curvature.
The geometry of the velodrome constitutes a special case: the transition curve is designed to establish a continuous change in curvature from a straight line to the arc of a circle of radius~$R$, or vice versa.
We use polynomial functions to model the transition, which requires solving the general Hermite interpolation problem.
In other words, we calculate curvature polynomial $\kappa(s) = f(s)$ that satisfies 
\begin{equation*}
	\kappa^{(j)}(s_i) = f^{(j)}(s_i),\quad
	j=0,1,\dots,\ell_i,\quad
	i=0,1,\dots,m,
\end{equation*}
where $\{s_i\}$ are the $(m+1)$ distinct points within a specified interval $[a,b]$ listed in ascending order $a\le s_0<s_1<\cdots<s_m\le b$, and $\ell_i$ is the highest order of the derivative at each $s_i$.
The solving procedure is detailed in~\citet[Section~5.5]{Powell1981}, which we summarize as follows.

\def\arraystretch{0.75}%
\begin{table}
	\centering
	\begin{tabular}{c*{5}{c}}
		$\xi_i$ & $f[\xi_i]$ & Order 1 & Order 2 & $\cdots$ & Order $n$\\
		\toprule 
		$\xi_0$ & $f[\xi_0]$ & \multirow{3}{*}{$f[\xi_0,\xi_1]$} & \multirow{5}{*}{$f[\xi_0,\xi_1,\xi_2]$} & \multirow{7}{*}{$\ddots$} & \multirow{9}{*}{$f[\xi_0,\xi_1,\dots,\xi_n]$} \\
		& \\
		$\xi_1$ & $f[\xi_1]$ & \multirow{3}{*}{$f[\xi_1,\xi_2]$} & \multirow{5}{*}{$\vdots$} & \multirow{7}{*}{$\iddots$}\\
		& \\
		$\xi_2$ & $f[\xi_2]$ & \multirow{3}{*}{$\vdots$} & \multirow{5}{*}{$f[\xi_{n-2},\xi_{n-1},\xi_n]$} \\
		& \\
		$\vdots$ & $\vdots$ & \multirow{3}{*}{$f[\xi_{n-1},\xi_n]$}\\
		& \\
		$\xi_n$ & $f[\xi_n]$ & \\
		\bottomrule
	\end{tabular}
	\caption{Divided differences table (Adapted from~\citet[Section~5.3]{Powell1981})}
	\label{tab:DivDiff}
\end{table}
\def\arraystretch{1}%

The procedure requires the application of Newton's interpolation method to an organized data set of function and derivative values $\{f^{(j)}(s_i)\}$ such that 
{\small
\begin{equation*}
	\{f^{(0)}(s_0),f^{(1)}(s_0),\dots,f^{(\ell_0)}(s_0),f^{(0)}(s_1),\dots,f^{(\ell_1)}(s_1),\dots,f^{(0)}(s_m),\dots,f^{(\ell_m)}(s_m)\}.
\end{equation*}
}%
The indexing of this set results in a grouping of $(\ell_i+1)$ repeated $s_i$ points, which we reindex as $\{f(\xi_k)\}$, where $\{\xi_k;\,k=0,1,\dots,n\}$ and $n = \sum_{i=0}^m(\ell_i+1)$.
Thus, the interpolating polynomial is
\begin{equation*}
	\kappa(s) = f(\xi_0) + \sum\limits_{j=1}^{n-1} f[\xi_0,\dots,\xi_j]\left(\prod\limits_{k=0}^{j-1}(x-\xi_k)\right).
\end{equation*}
The coefficients are divided differences, displayed in Table~\ref{tab:DivDiff}, and are obtained using recurrence relation
\begin{equation*}
	f[\xi_j,\xi_{j+1},\dots,\xi_{j+k+1}] = \frac{f[\xi_{j+1},\dots,\xi_{j+k+1}]-f[\xi_j,\dots,\xi_{j+k}]}{\xi_{j+k+1} - \xi_j},
\end{equation*}
where $f[\xi_j] = f(\xi_j)$ and, in the case of $\xi_{j+k+1}-\xi_j = 0$, we use
\begin{equation*}
	f[\xi_j,\xi_{j+1},\dots,\xi_{j+k+1}] = \frac{f^{(k+1)}(\xi_j)}{(k+1)!}.
\end{equation*}

Let us demonstrate the procedure for three cases of $\ell_0$ and $\ell_1$.
First, the simplest case of $\ell_0=\ell_1 = 0$ yields an interpolation of $(n=2)$ distinct points, $f(s_0) = \kappa_0$ and $f(s_L) = \kappa_L$, without considering derivatives, where, $\kappa_L$ is the curvature at the end of the interval, $s_L$.
We set $\{\xi_i\} = \{s_0,s_L\}$, and $\{f(\xi_i)\} = \{\kappa_0,\kappa_L\}$.
Using the divided differences
\begin{equation*}
	f[\xi_0] = \kappa_0,\quad
	f[\xi_1] = \kappa_L,
	\quad\text{and}\quad
	f[\xi_0,\xi_1] = \frac{f[\xi_1]-f[\xi_0]}{\xi_1-\xi_0} = \frac{\kappa_L-\kappa_0}{L},
\end{equation*}
the linear interpolation polynomial is
\begin{equation}
	\label{eq:kappa(s)_lin}
	\kappa(s) = f(\xi_0) + f[\xi_0,\xi_1]\,(s-\xi_0) =  \kappa_0 + \left(\kappa_L-\kappa_0\right)\left(\frac{s-s_0}{L}\right).
\end{equation}

Second, setting $\ell_0=\ell_1=1$ yields an interpolation of both the function values, $f(s_0)$ and $f(s_L)$, and their first derivatives, $f'(s_0)=f'(s_L)=0$.
In this case, the $(n=4)$ interpolation points are $\{\xi_i\} = \{s_0,s_0,s_L,s_L\}$ and the values are $\{f(\xi_i)\} = \{f(s_0),f'(s_0),f(s_L),f'(s_L)\} = \{\kappa_0,0,\kappa_L,0\}$.
Using the divided differences presented in Table~\ref{tab:ell0=ell1=1}, the cubic interpolating polynomial is
\begin{equation}
	\label{eq:kappa(s)_cub}
	\kappa(s) = f(\xi_0) + \sum\limits_{j=1}^{3} f[\xi_0,\dots,\xi_j]\left(\,\prod\limits_{k=0}^{j-1}(x-\xi_k)\right) \\
	=
	\kappa_0 + (\kappa_L-\kappa_0)\left(\frac{(s-s_0)^2}{L^2} - \frac{2\,(s-s_0)^2(s-s_L)}{L^3}\right).
\end{equation}
Third, selecting $\ell_0=\ell_1=2$ yields the interpolation of the function values and up to their second derivatives, $f''(s_0) = f''(s_L) = 0$.
The resulting quintic interpolation polynomial is
\begin{equation}
	\label{eq:kappa(s)_quin}
	\kappa(s) = \kappa_0 + (\kappa_L-\kappa_0)\left(\frac{(s-s_0)^3}{L^3} - \frac{3\,(s-s_0)^3(s-s_L)}{L^4} + \frac{6\,(s-s_0)^3(s-s_L)^2}{L^5}\right).
\end{equation}

\def\arraystretch{0.75}%
\begin{table}
	\centering
	\begin{tabular}{r*{4}{r}}
		$\xi_i$ & $f[\xi_i]$ & Order 1 & Order 2 & Order 3 \\
		\toprule
		$s_0$ & $\kappa_0$ & \multirow{3}{*}{0} & \multirow{5}{*}{$(\kappa_L-\kappa_0)/L^2$} & \multirow{7}{*}{$-2\,(\kappa_L-\kappa_0)/L^3$} \\
		& \\
		$s_0$ & 0 & \multirow{3}{*}{$(\kappa_L-\kappa_0)/L$} & \multirow{5}{*}{$-(\kappa_L-\kappa_0)/L^2$}\\
		& \\
		$s_L$ & $\kappa_L$ & \multirow{3}{*}{0} \\
		& \\
		$s_L$ & 0 \\
		\bottomrule
	\end{tabular}
	\caption{Divided difference table for $\ell_0=\ell_1=1$}
	\label{tab:ell0=ell1=1}
\end{table}
\def\arraystretch{1}%

Now, depending on the values of $\kappa_0$ and $\kappa_L$, polynomial expressions~\eqref{eq:kappa(s)_lin}--\eqref{eq:kappa(s)_quin} will change accordingly.
In the case of ascending curvature from a straight line, $\kappa_0 = 0$, to the arc of a circle, $\kappa_L = 1/R$, the respective expressions become
\begin{subequations}
	\label{eq:kappa(s)_asc}
	\begin{equation}
		\kappa(s) = \frac{s}{R\,L},\quad
		\kappa(s) = \frac{1}{R}\left(\frac{3\,s^2}{L^2} - \frac{2\,s^3}{L^3}\right),\quad
		\kappa(s) = \frac{1}{R}\left(\frac{10\,s^3}{L^3} - \frac{15\,s^4}{L^4} + \frac{6\,s^5}{L^5}\right),
		\tag{\theequation a--c}
	\end{equation}
\end{subequations}
where curvature~(\ref{eq:kappa(s)_asc}a) gives rise to the so-called Euler spiral,~(\ref{eq:kappa(s)_asc}b) is the so-called Bloss curvature~(\citet{Bloss1936}, as cited in~\citet{Kufver1997}), and~(\ref{eq:kappa(s)_asc}c) is the so-called Watorek curvature~(\citet{Watorek1907}, as cited in~\citet{Kufver1997}).

Other combinations of $\ell_i$ and interpolation data points lead to various expressions of curvature.
The approach detailed in this section may be used to determine various curvature profiles, depending on the modelling requirements.

\section{Assembly procedure}
In this section, we demonstrate the velodrome surface assembly procedure. 
We restrict our attention to Category~1 velodromes for the World Championships and Olympic Games.
In accordance with Section~\ref{sec:UCI}, we recall that the design must satisfy
\begin{equation*}
	L = 250\,{\rm m},\quad
	R\in[19\,{\rm m},25\,{\rm m}],\quad
	w\in[7\,{\rm m},8\,{\rm m}],\quad
	w_B \geq w/10,\quad
	w_{SZ} + w_B \geq 4\,{\rm m}.
\end{equation*}
Also, Article~3.6.067 requires the form of the inner edge of the track to consists of two parallel straight lines are connected by two curves, and Article~3.6.069 determines that length of the track is measured 20\,cm above the inner edge of the track.

\begin{table}
	\centering
	\begin{subfigure}[b]{0.48\textwidth}
		\centering
		\begin{tabular}[t]{ccc}
			$S_i$ & rate of change & values \\
			\toprule
			$S_1$ & constant &$12^\circ$ \\
			$S_2$ & increase & sinusoid \\
			$S_3$ & constant & $45^\circ$ \\
			\bottomrule
		\end{tabular}
		\caption{$\phi(s)$}
		\label{tab:symVD_phi}
	\end{subfigure}
	\quad
	\begin{subfigure}[b]{0.48\textwidth}
		\centering
		\begin{tabular}[t]{ccc}
			$S_i$ & rate of change & values \\
			\toprule
			$S_1$ & constant & 0 \\
			$S_2$ & increase & linear \\
			$S_3$ & constant & $1/R$ \\
			\bottomrule
		\end{tabular}
		\caption{$\kappa(s)$}
		\label{tab:symVD_kappa}
	\end{subfigure}
	\caption{%
		Symmetric banking and curvature profiles (only the first-quadrant design is shown). 
		The banking angle~(\subref{tab:symVD_phi}) and curvature~(\subref{tab:symVD_kappa}) changes are restricted to the transition curve, with a turn radius of $R = R_R = R_L = 21.5$\,m.}
	\label{tab:symVD_phi_kappa}
\end{table}

To comply with the form regulation, we use
\begin{equation*}
	\kappa_{12}(s) = \kappa_1(s) = \kappa_6(s) = \kappa_7(s) = 0,\quad
	\kappa_3(s) = \kappa_4(s) = \frac{1}{R_R},\quad
	\kappa_9(s) = \kappa_{10}(s) = \frac{1}{R_L},
\end{equation*}
to index the straight-line and circular-turn segments; herein, $R_R$ and $R_L$ are the left- and right-turn radii.
Necessarily, the remaining segments are the transition curves.
Initial values ${\boldsymbol\alpha}(s_0) = O$ and $\theta_1 = 0$ fix the home straightaway parallel to the abscissa, with the initial directrix coordinate at the origin, $O$.
For a parallel back straightaway, but oriented in the opposite direction, we require $\theta_6 = \theta_1 + \pi = \pi$.
Then, closure conditions~\eqref{eq:theta_{12}} and~\eqref{eq:r_{12}} simplify to $\theta_{12} = \theta_1 + 2\pi = 2\pi$ and ${\boldsymbol\alpha}(s_{12}) = {\boldsymbol\alpha}(s_{0}) = O$.
To comply with the length regulation, we set lap length~\eqref{eq:L_L} to $L_L = L$.
Therefore, up to the specification of $\phi(s)$ and $\kappa(s)$, the segment lengths, $\boldsymbol l$, of a Category~1 track must satisfy
\begin{equation}
	\label{eq:h(l)=0}
	{\boldsymbol h}({\boldsymbol l}) = [L_L - L,\, \theta_6 - \pi,\, \theta_{12} - 2\pi,\, {\boldsymbol\alpha}(s_{12}) - {\boldsymbol\alpha}(s_{0})]^t.
\end{equation}

\begin{table}
	\centering
	\begin{subfigure}[b]{0.48\textwidth}
		\centering
		\begin{tabular}[t]{cccc}
			$S_i$ & form & \multicolumn{2}{c}{ranges} \\
			\toprule
			$S_1\cup S_2\cup S_3$ & linear & $12^\circ$ & $40^\circ$  \\
			$S_4\cup S_5\cup S_6$ & sinusoid & $40^\circ$ & $17^\circ$  \\
			$S_7$ & constant & \multicolumn{2}{c}{$17^\circ$} \\
			$S_8\cup S_9$ & sinusoid & $17^\circ$ & $46^\circ$  \\
			$S_{10}\cup S_{11}$ & cubic & $46^\circ$ & $12^\circ$  \\
			$S_{12}$ & constant & \multicolumn{2}{c}{$12^\circ$} \\
			\bottomrule
		\end{tabular}
		\caption{$\phi(s)$}
		\label{tab:asymVD_phi}
	\end{subfigure}
	\quad
	\begin{subfigure}[b]{0.48\textwidth}
		\centering
		\begin{tabular}[t]{cccc}
			$S_i$ & rate of change & values \\
			\toprule
			$S_{12}\cup S_1$ & constant & 0 \\
			$S_2$ & increase & linear \\
			$S_3\cup S_4$ & constant & $1/R_R$ \\
			$S_5$ & decrease & quintic \\
			$S_6\cup S_7$ & constant & 0 \\
			$S_8$ & increase & cubic \\
			$S_9\cup S_{10}$ & constant & $1/R_L$ \\
			$S_{11}$ & decrease & linear \\
			\bottomrule
		\end{tabular}
		\caption{$\kappa(s)$}
		\label{tab:asymVD_kappa}
	\end{subfigure}
	\caption{%
		Asymmetric banking and curvature profiles. 
		The banking angles~(\subref{tab:asymVD_phi}) are inconsistent between the home and back straightaways as well as the left- and right-circular turns, and change according to different forms across multiple segments.
		The curvature~(\subref{tab:asymVD_kappa}) varies according to different polynomial functions with inconsistent turn radii, $R_R = 23$\,m and $R_L = 20$\,m.}
	\label{tab:asymVD_phi_kappa}
\end{table}

Since there is a surplus of variables in comparison to only four constraints, in tandem with the functional freedom of $\phi(s)$ and $\kappa(s)$, the design problem is underdetermined and, as such, there are an infinite number of solutions.
The underdetermination is less severe for symmetric designs due to the reflection symmetry across the $x$ and $y$ axes into the remaining quadrants.
However, underdetermination increases for asymmetric designs as each quadrant consists of its own unique design features.
In both cases, we manage the freedom of design by specifying~$\phi(s)$ and~$\kappa(s)$ prior to using numerical optimization to solve for the segment lengths that satisfy~${\boldsymbol h}$.
Since the lengths determine the limits of lap-length integration, we choose the Nelder-Mead algorithm~(e.g., \citet[Chapter~4.3]{Cavazzuti2013}), which is a local derivative-free simplex technique that only uses function values.
We use the algorithm to minimize the residual sum of squares of~${\boldsymbol h}$, which is tantamount to obtaining a least-squares solution to the design problem. 

\begin{figure}
	\centering
	\begin{subfigure}[b]{\textwidth}
		\includegraphics[width=0.485\textwidth]{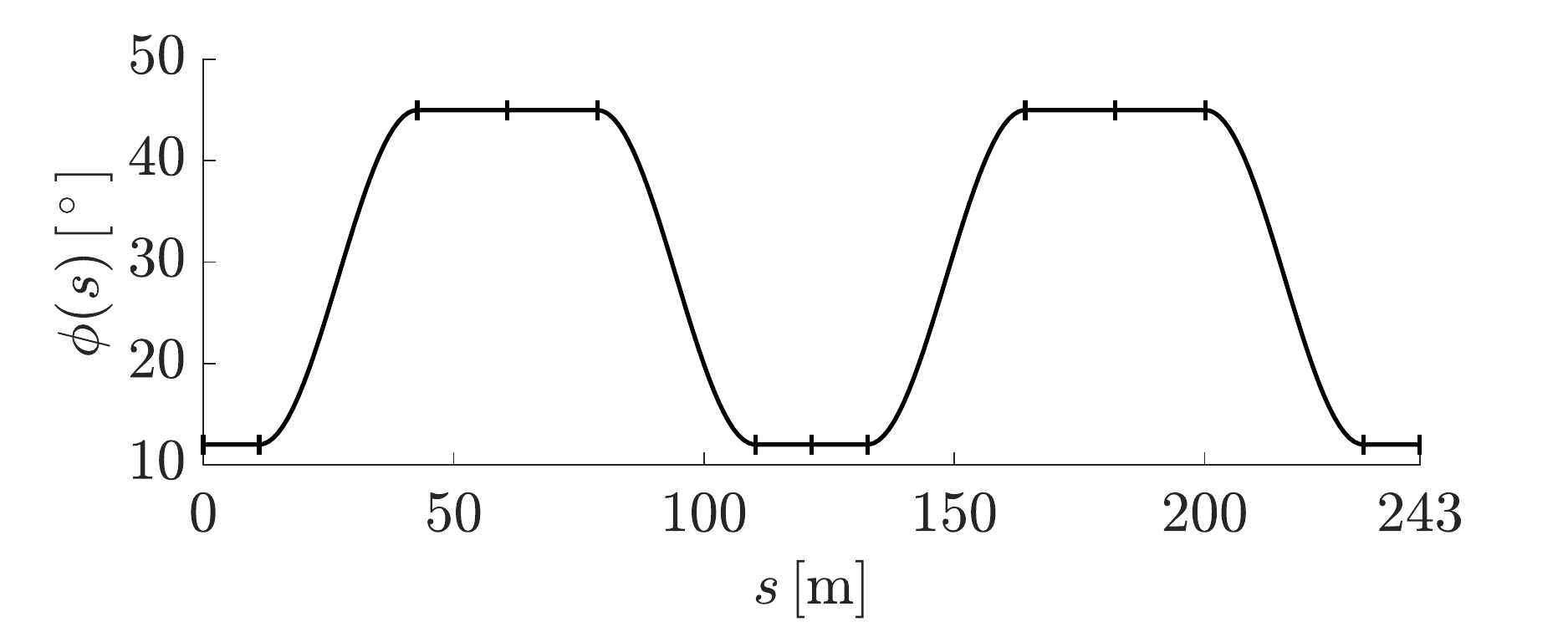}
		\hfill
		\includegraphics[width=0.485\textwidth]{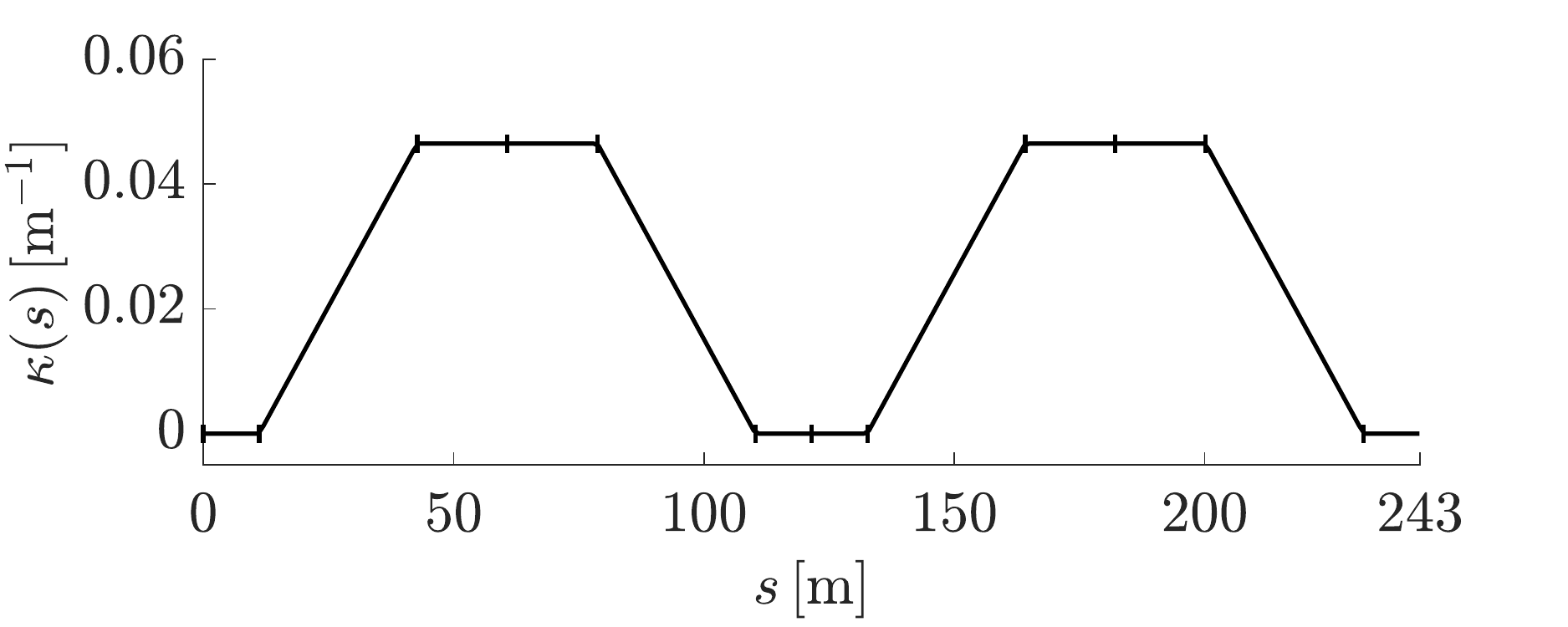}
		\caption{}
		\label{fig:symVD_phi_kappa}
	\end{subfigure}
	\\
	\begin{subfigure}[b]{\textwidth}
		\includegraphics[width=0.485\textwidth]{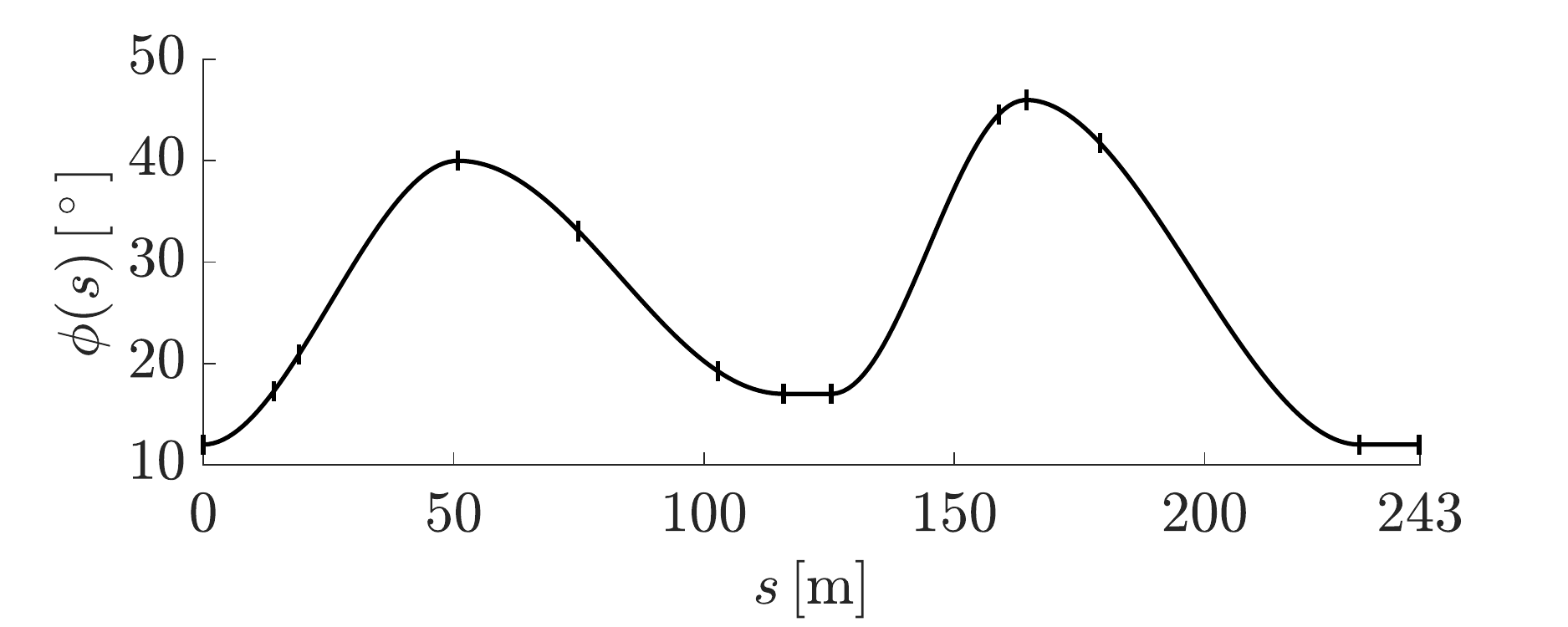}
		\hfill
		\includegraphics[width=0.485\textwidth]{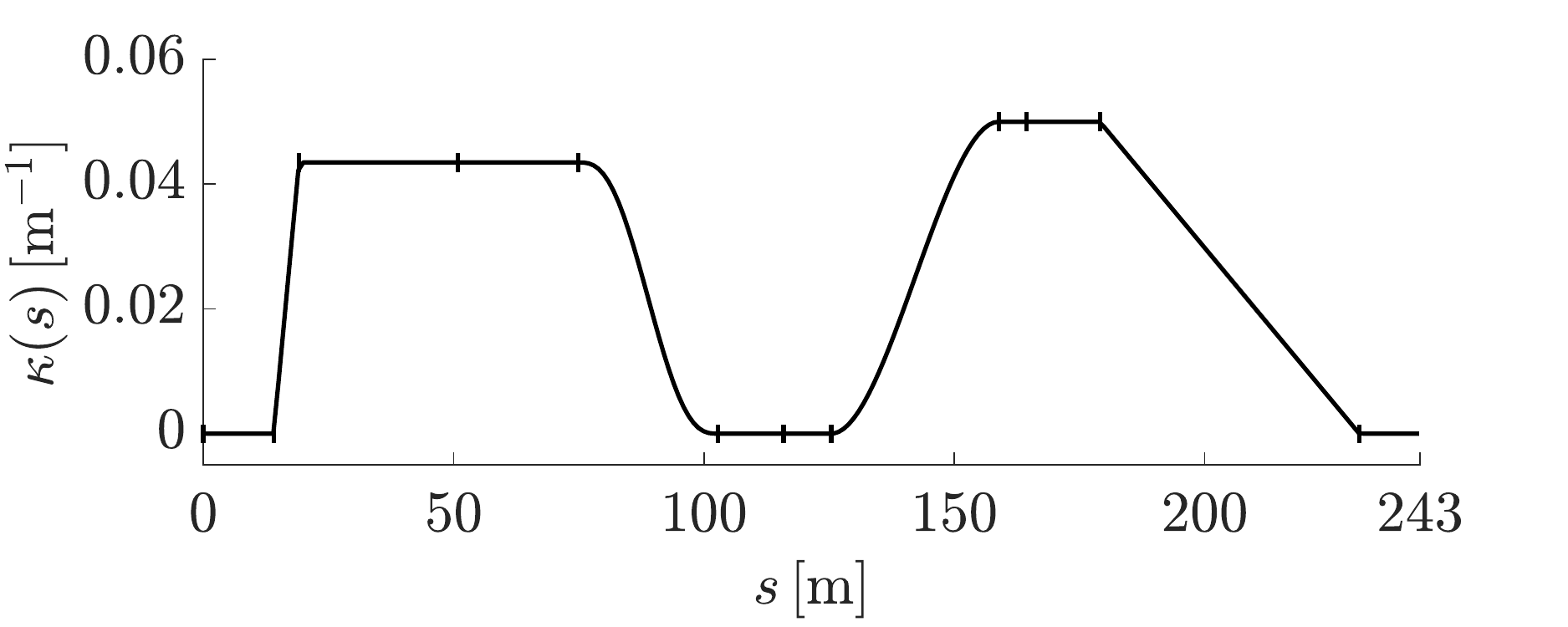}
		\caption{}
		\label{fig:asymVD_phi_kappa}
	\end{subfigure}
	\caption{%
	Banking (left-hand side) and curvature (right-hand side) profiles for symmetric~(\subref{fig:symVD_phi_kappa}) and asymmetric~(\subref{fig:asymVD_phi_kappa}) velodrome designs: abscissa indicates directrix arc length; perpendicular ticks superimposed on curve indicate segment boundaries.}
	\label{fig:phi_kappa}
\end{figure}

\begin{figure}
	\centering
	\begin{subfigure}[b]{0.495\textwidth}
		\includegraphics[width=\textwidth]{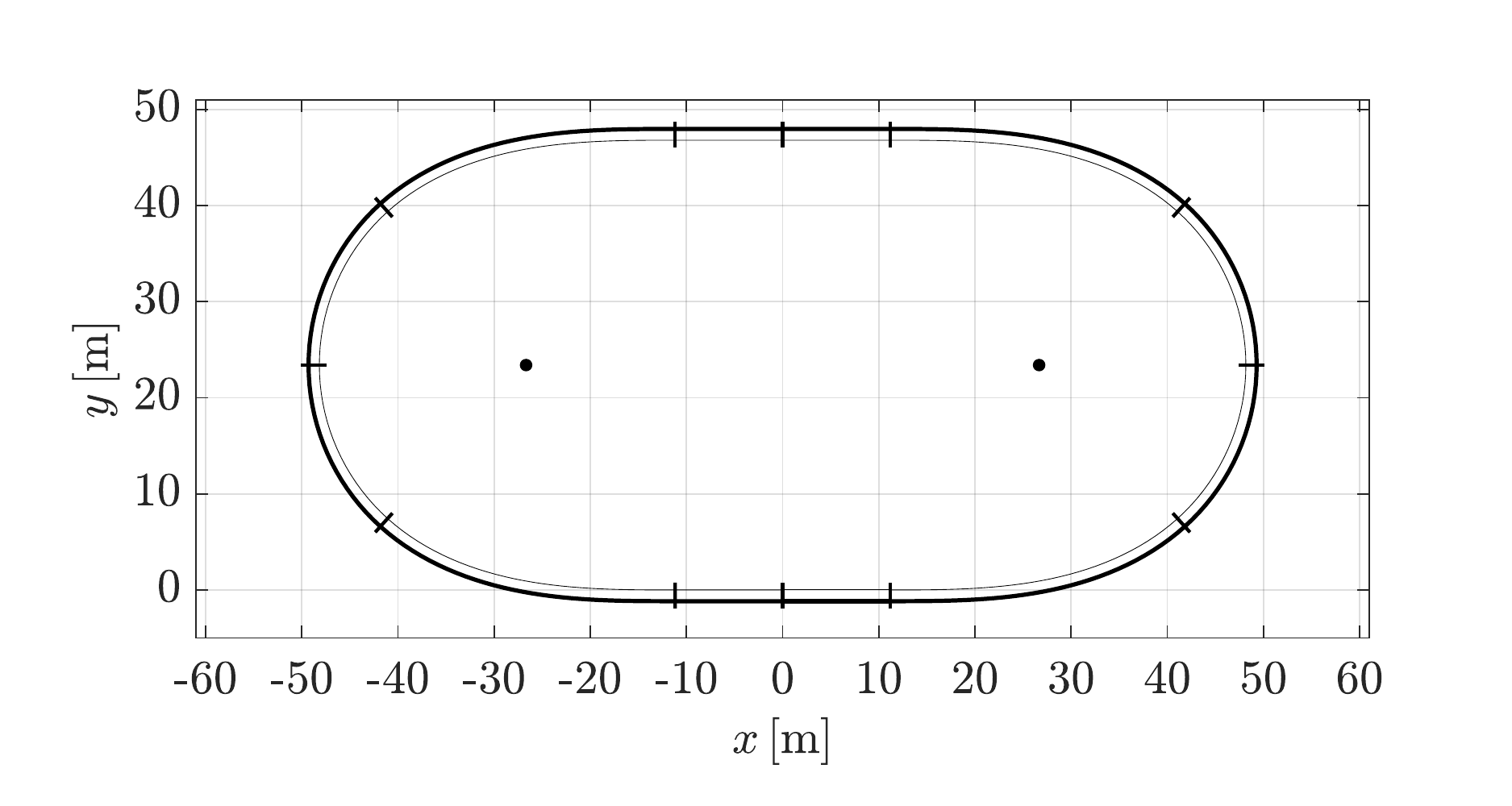}
		\caption{}
		\label{fig:symVD_plan}
	\end{subfigure}
	\hfill
	\begin{subfigure}[b]{0.495\textwidth}
		\includegraphics[width=\textwidth]{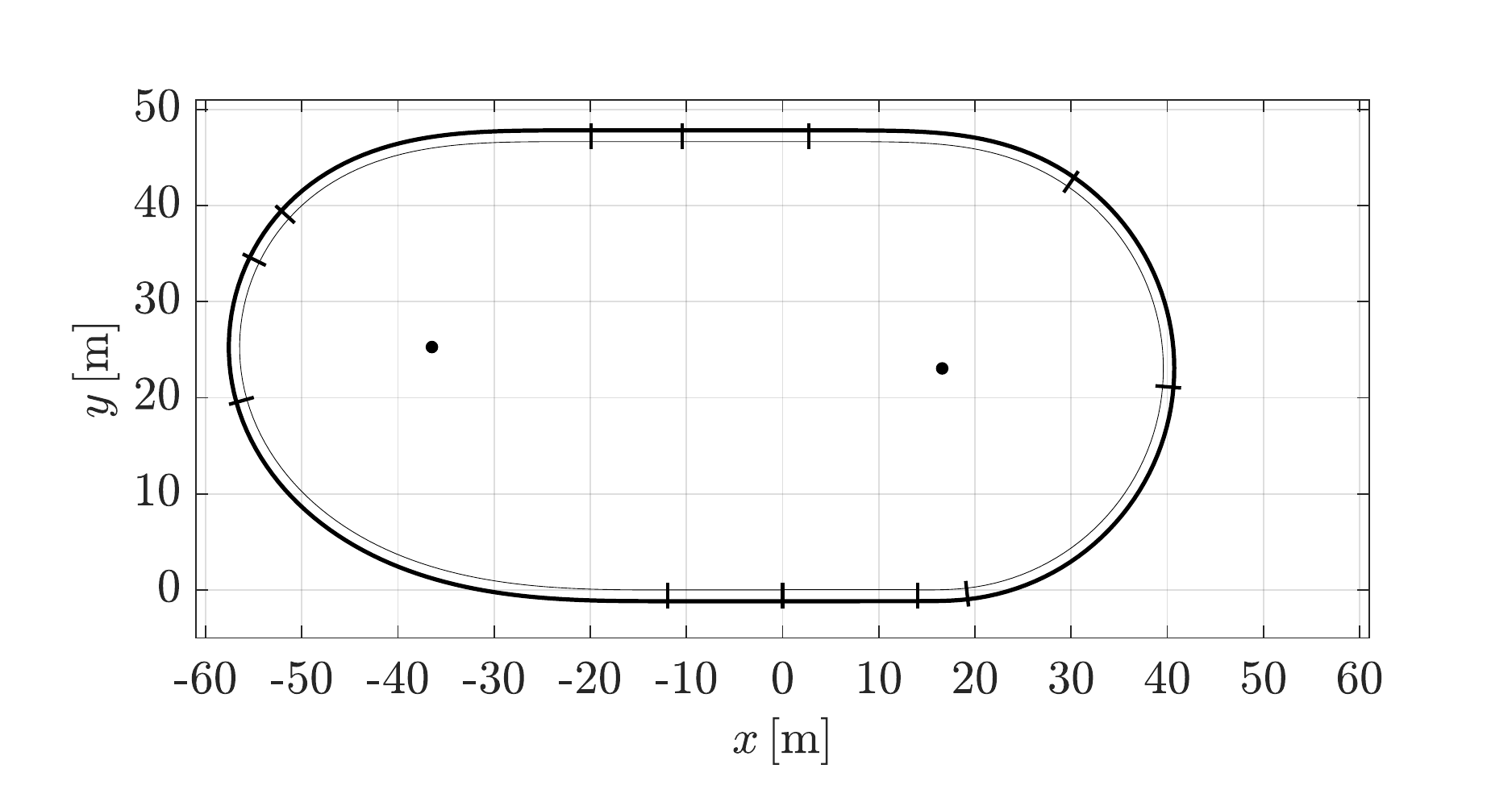}
		\caption{}
		\label{fig:asymVD_plan}
	\end{subfigure}
	\caption{%
		Plan view of symmetric~(\subref{fig:symVD_plan}) and asymmetric~(\subref{fig:asymVD_plan}) velodrome designs:
		thin inner line is the directrix; bold outer line is the measuring line; perpendicular ticks superimposed on curves indicate segment boundaries; dots indicate circular turn origins.}
	\label{fig:plan}
\end{figure}

\begin{figure}
	\begin{subfigure}[b]{\textwidth}
		\centering
		\includegraphics[width=0.85\textwidth]{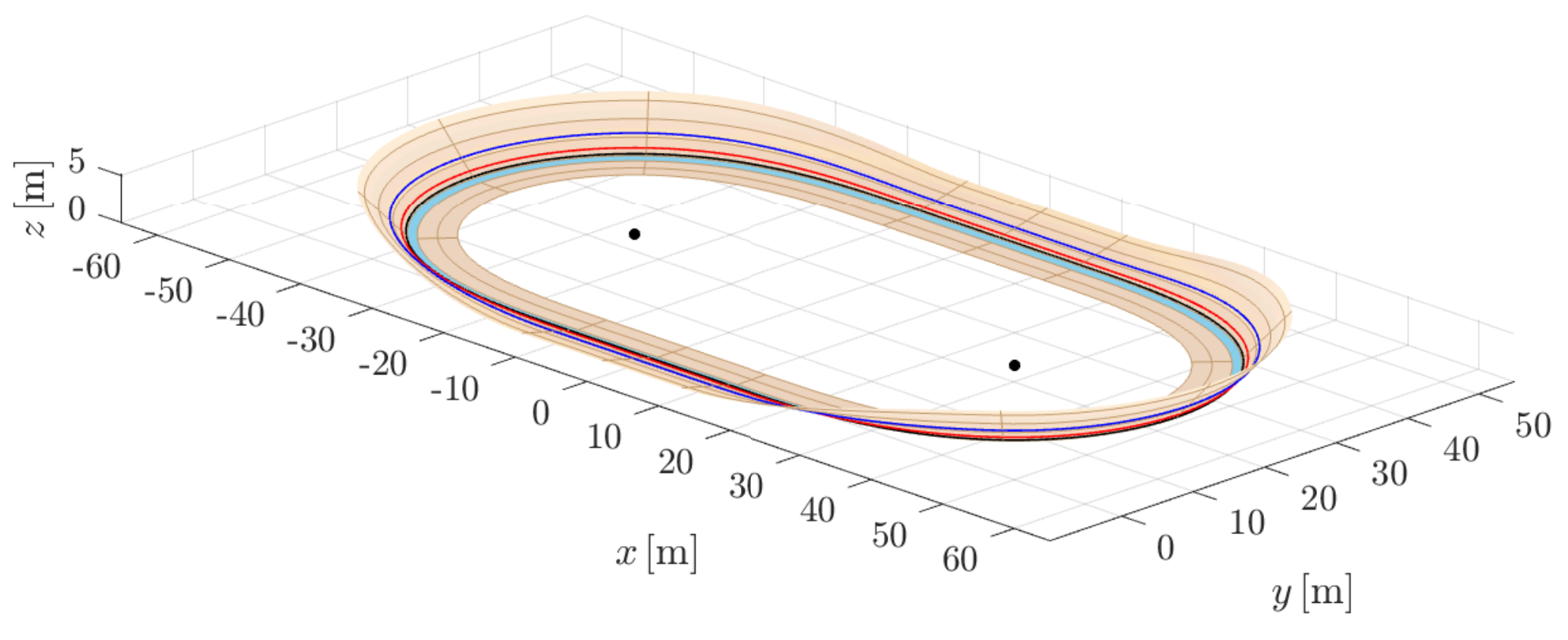}
		\caption{}
		\label{fig:symVD_3D}
	\end{subfigure}
	\\
	\begin{subfigure}[b]{\textwidth}
		\centering
		\includegraphics[width=0.85\textwidth]{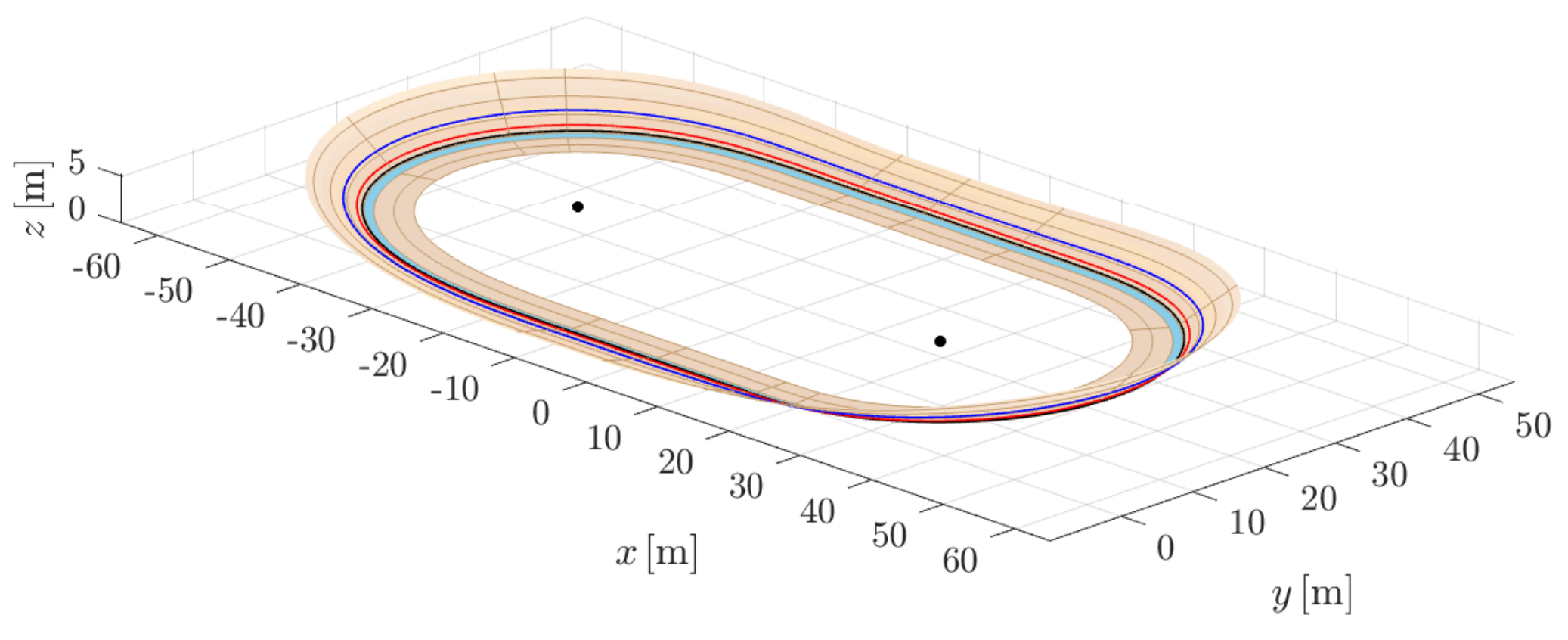}
		\caption{}
		\label{fig:asymVD_3D}
	\end{subfigure}
	\caption{%
		3D renderings of symmetric (\subref{fig:symVD_3D}) and asymmetric~(\subref{fig:asymVD_3D}) velodromes: axis orientation (view of back straightaway) consistent with Figure~\ref{fig:plan}; blue band separates safety zone and track: brown lines perpendicular to directrix indicate segment boundaries; black, red, and blue lines offset to directrix indicate measuring, sprinters', and stayers' lines; dots indicate circular turn origins.}
	\label{fig:3D}
\end{figure}

Let us use this approach to compare a symmetric and asymmetric velodrome design.
In both cases, we set
\begin{equation*}
	w = 7\,{\rm m},\quad
	w_B = 1\,{\rm m},\quad
	w_{SZ} = 4\,{\rm m},\quad\text{and}\quad
	\phi_B = 12^\circ.
\end{equation*}
For the symmetric design, we specify $\phi(s)$ and $\kappa(s)$ in Table~\ref{tab:symVD_phi_kappa} and, then, we use the algorithm to obtain a least-squares solution, for which the pertinent directrix segment lengths,
\begin{equation*}
	l_1 = 11.18\,{\rm m},\quad
	l_2 = 31.56\,{\rm m},\quad
	l_3 = 17.99\,{\rm m},
\end{equation*}
result in a directrix period of $s_{12} = 242.92\,{\rm m} < L$.
Using the segments, we plot $\phi(s)$ and $\kappa(s)$ in Figure~\ref{fig:phi_kappa}(\subref{fig:symVD_phi_kappa}) and, in Figure~\ref{fig:plan}(\subref{fig:symVD_plan}), we illustrate the directrix and measuring-line symmetry in a plan-view plot.
The lengths of the segments, as measured at ruling $v_M$ on the velodrome surface, are
\begin{equation*}
	L_{{\bf r}_M}(s_0,s_1) = 11.18\,{\rm m},\quad
	L_{{\bf r}_M}(s_1,s_2) = 32.39\,{\rm m},\quad
	L_{{\bf r}_M}(s_2,s_3) = 18.93\,{\rm m},
\end{equation*}
which may be used to verify the lap length equals $L$.
We provide a 3D computer rendering of the symmetric velodrome design in Figure~\ref{fig:3D}(\subref{fig:symVD_3D}).

While a symmetric design may be sufficient for modelling purposes, a general formulation must accommodate design asymmetry.
To demonstrate the versatility of our approach, let us model the following contrived, yet UCI-compliant, design: different relative lengths and banking of the home and back straightaways, inconsistent turn radii and banking along the bends, and mismatching transition curves in each quadrant.
We proceed by specifying $\phi(s)$ and $\kappa(s)$ in Table~\ref{tab:asymVD_phi_kappa}, then deploying the optimization algorithm to obtain 
\begin{equation*}
	{\boldsymbol l} = [14.06, 5.07, 31.71, 24.12, 27.78, 13.18, 9.52, 33.42, 5.56, 14.65, 51.81, 11.94]^t,
\end{equation*}
which yield a directrix period of $s_{12} = 242.83$\,m. 
The plots of~$\phi(s)$ and~$\kappa(s)$ in Figure~\ref{fig:phi_kappa}(\subref{fig:asymVD_phi_kappa}) demonstrate the asymmetric design, which feature rapid increases in banking and curvature at the entrances of the turns and gradual decreases at the exits.
In Figure~\ref{fig:plan}(\subref{fig:asymVD_plan}), the plan view illustrates the misalignment of the centres of the circular turns and the straightaways.
Also, since the ends of intervals $S_3$ and $S_9$ are misaligned with the apexes of the turns, the banking angle begins to decline on the entrances of the turns.
The illustration of these asymmetric features is shown in the 3D computer rendering of Figure~\ref{fig:3D}(\subref{fig:asymVD_3D}).
Therein, along the measuring line ruling, the segment lengths are
\begin{equation*}
	[14.06, 5.20, 33.29, 25.31, 28.48, 13.18, 9.52, 34.37, 5.88, 15.47, 53.30, 11.94]^t,
\end{equation*}
which may be used to verify the lap length $L$.

\section{Conclusion}
This article details a general approach to velodrome track design.
The mathematical model is developed from first principles, rooted in differential geometry, and complies with the regulations mandated by the UCI.
The velodrome is presented as a ruled surface, for which the geometric properties of its directrix determine the form of the track and its director curve generates the surface.
For the purpose of a general formulation, we construct the directrix on a per-segment basis, each as a function of the banking and curvature along its length.
Since the design problem is underdetermined, there are an infinite number of solutions.
To obtain particular solutions, we specify {\it a priori} the desired banking and curvature profiles and, then, use numerical optimization to find a set of directrix segment lengths that satisfy a system of equations.
We present a symmetric and asymmetric velodrome design and compare the results to demonstrate the versatility of our novel approach.
The banking and curvature of the former are restricted to vary in a sinusoidal and linear manner, respectively, along its transition curves, whereas the features of the latter are more sophisticated and different in every quadrant of the velodrome.

\section*{Acknowledgements}
We wish to acknowledge the following professors, without whom this article would not be possible:
\mbox{Michael A. Slawinski}, for his unwavering mentorship and for fostering an enthusiasm toward cycling-related mathematical modelling;
\mbox{Len Bos}, for sharing his mathematical rigour, to which I strive continually to emulate; and
\mbox{Ivan Booth}, for sparking an interest in differential geometry and its infinite applications.

\bibliographystyle{apa}
\bibliography{S_VD.bib}
\end{document}